# Astro2020 Science White Paper

# Searching for Exosatellites Orbiting L and T Dwarfs: Connecting Planet Formation to Moon Formation and Finding New Temperate Worlds

**Thematic Areas:** ☒ Planetary Systems    ☐ Star and Planet Formation
☐ Formation and Evolution of Compact Objects    ☐ Cosmology and Fundamental Physics
☒ Stars and Stellar Evolution    ☐ Resolved Stellar Populations and their Environments
☐ Galaxy Evolution    ☒ Multi-Messenger Astronomy and Astrophysics

**Principal Author:**
Name: Philip S. Muirhead
Institution: Boston University
Email: philipm@bu.edu
Phone: 1-617-353-6553

**Co-authors:**
Julie N. Skinner (Smith College, Northampton, MA, jnskinner@smith.edu);
Jacqueline Radigan (Utah Valley University, Orem, UT, radigan@uvu.edu);
Amaury Triaud (University of Birmingham, Birmingham, UK, a.triaud@bham.ac.uk);
Christopher Theissen (University of California, San Diego, CA, ctheissen@ucsd.edu);
Daniella Bardalez Gagliuffi (American Museum of Natural History, New York, NY, dbardalezgagliuffi@amnh.org);
Patrick Tamburo (Boston University, Boston, MA, tamburop@bu.edu);
Adam Burgasser (University of California, San Diego, CA, aburgasser@ucsd.edu);
Jacqueline Faherty (American Museum of Natural History, New York, NY, jfaherty@amnh.org);
Denise Stephens (Brigham Young University, Provo, UT, denise_stephens@byu.edu)

**Abstract**: L-type and T-type dwarfs span the boundaries between main-sequence stars, brown dwarfs, and planetary-mass objects. For these reasons, L and T dwarfs are the perfect laboratories for exploring the relationship between planet formation and moon formation, and evidence suggests they may be swarming with close-in rocky satellites, though none have been found to date. The discovery of satellites orbiting L or T dwarfs will have transformative implications for the nature of planets, moons and even life in the Universe. These transiting satellites will be prime targets for characterization with NASA's *James Webb Space Telescope*. In this white paper, we discuss the scientific motivations behind searching for transiting satellites orbiting L and T dwarfs and argue that robotizing current 1-2-meter US optical/infrared (O/IR) facilities and equipping them with recently developed low-cost infrared imagers will enable these discoveries in the next decade. Furthermore, robotizing the 1-2-meter O/IR fleet is highly synergistic with rapid follow-up of transient and multi-messenger events.

## Introduction: L and T Dwarfs as Transition Objects

L-type and T-type dwarfs span the boundaries between main-sequence stars, brown dwarfs, and planetary-mass objects (Kirkpatrick et al. 1999; Burgasser et al. 2006). The L and T spectral types describe dwarf objects with effective temperatures that are cooler than M type, the latest spectral type in the Harvard classification system (Cannon and Pickering 1925). They have masses that range between 0.005 and 0.080 solar, and due to the effects of electron degeneracy in their cores, they all have radii roughly equal to that of Jupiter. The L0, L1 and L2-type dwarfs can be either hydrogen-fusing stars or young brown dwarfs (e.g., Dieterich et al. 2014; Baraffe et al. 2015). The L3 and later (cooler) types are either brown dwarfs, having undergone a brief period of deuterium fusion (e.g., Baraffe et al. 2002), or planetary-mass objects that never experienced any fusion processes, radiating heat from their formation for the remainder of their lives (Burrows et al. 1997). The ambiguity as to whether a given L or T dwarf is a star, a brown dwarf or a planetary-mass object can be resolved by estimating its age. Figure 1 shows evolutionary models for these objects from Burrows et al. (1997), illustrating the overlap between stars, brown dwarfs and planets at a given L or T spectral type.

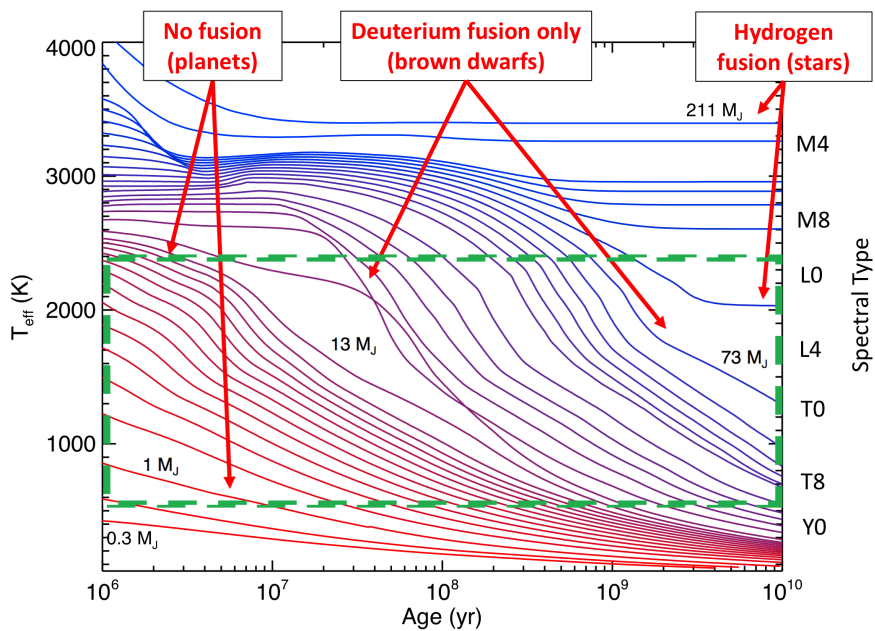

**Figure 1:** Evolutionary tracks for low-mass stars, brown dwarfs and planetary-mass objects, showing effective temperature ($T_{eff}$) and spectral type versus age, color-coded according to initial mass (adapted from Burrows et al. 1997). Each line represents a single object. L and T dwarfs (dashed green box) can be either stars, brown dwarfs or planets depending on their ages.

By spanning the boundaries between stars, brown dwarfs and planets, L and T dwarfs are the perfect laboratories for testing the relationship between planet formation and moon formation. That is to say, they are the key to a general theory of *satellite* formation, and evidence suggests they should be swarming with close-in Earth-radius rocky satellites: Thanks to NASA's *Kepler* Mission, we now know that M dwarf stars host at least two short-period, Earth-radius exoplanets per star (e.g., Dressing & Charbonneau 2015) with a significant fraction in co-planar multi-planet systems (Muirhead et al. 2015; Ballard & Johnson 2016). On the planet side, it is well known that every gas-giant planet in our own solar system hosts a system of moons. This suggests that the objects with masses *in between* stars and gas-giant planets, the L and T dwarfs, should also host systems of short-period satellites, including those that orbit in their hosts' habitable zones. Indeed, young precursors to L and T dwarfs show evidence of long-lived protoplanetary disks, with lifetimes of up to 50 Myr (Boucher et al. 2016), providing ripe conditions for forming satellites from coalescing metal, rock and ice within those disks (e.g., Luhman et al. 2005).

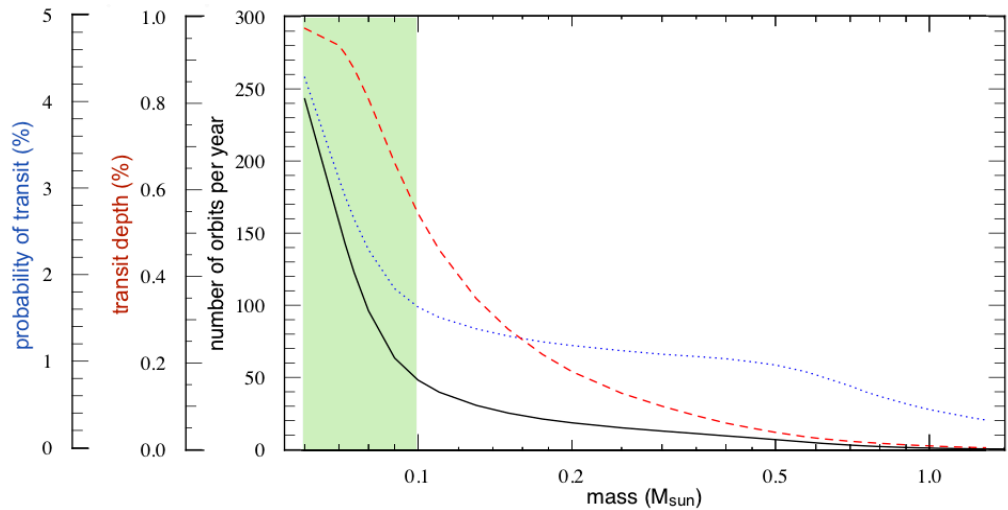

Figure 2 (from He et al. 2017, used with permission): For a given equilibrium temperature (here 255K, like Earth), the number of orbits (i.e., transits or occultations) per year (black), the transit depth (dashed red), and the probability of transit (dotted blue) as a function of the primary's mass. The green area corresponds to L and T dwarf stars. The natural evolution of stars and brown dwarfs will alter those curves with time; for brown dwarfs, the slopes get steeper (and more favorable) with time.

**Searching for Satellites Around L and T Dwarfs**

L and T dwarfs are exciting places to search for transiting rocky satellites. For a transiting planet or satellite, the fractional decrease in flux depends on the ratio of the sky-projected area of the satellite to that of the host. As all L and T dwarfs are roughly the size of Jupiter, an Earth-sized satellite introduces a 1% transit depth into the light curve of an L or T dwarf—a fractional transit depth achievable with ground-based telescopes. With bolometric luminosities between $10^{-5}$ and $10^{-4}$ that of the sun, the habitable-zones for L and T dwarfs are much closer-in than for Sun-like stars or even M-dwarf stars. Defined as the range of orbital distances where liquid water can be sustained on a rocky surface, the habitable zones for L and T dwarfs correspond to orbital periods of only a few days (Belu et al. 2013). Figure 2 illustrates the large transit depths, large probability of transits and sheer number of transits for a hypothetical system (He et al. 2017). For brown-dwarf and planetary-mass L and T dwarfs, the habitable zones will move closer and closer to the host as they perpetually radiate their heat into space and cool (see Figure 1). However, this also presents a unique opportunity in the hunt for habitable locations in the Universe: any satellite found with an orbital period of a few days around a cooling L or T dwarf will spend some time in their hosts' habitable zone (Bolmont 2017).

**Detailed Characterization of LT Exosatellites with the *James Webb Space Telescope***

In addition to the sheer importance of such a discovery, satellites transiting L and T dwarfs would provide golden opportunities for exoplanet characterization with the *James Webb Space Telescope*. The small size and low intrinsic luminosity of L and T dwarfs make them ideal targets for spaced-based characterization techniques, as does the short orbital period of the expected discoveries and frequent transit events. These characterization methods include secondary eclipse photometry to measure day/night contrasts and study exoplanet internal heating (e.g., Knutson et al. 2007; Croll et al. 2010, 2011) and transmission spectroscopy to measure the scale-heights of satellite atmospheres (e.g., Bean et al. 2010; Berta et al. 2012; Kreidberg et al. 2014). Add in the fact that all short-period satellites orbiting brown dwarf and planetary-mass L and T dwarfs spend some time in their hosts habitable zones, and characterization with *JWST* becomes truly transformative. A study by Belu et al. (2011) showed that the oxygen features, a biomarker, in a hypothetical transiting satellite would be easier to measure for M8 and later type stars with *JWST*.

Despite these advantages, L and T dwarfs have been largely ignored by transit and Doppler planet searches. In the case of the Doppler technique, L and T dwarfs are too intrinsically faint to detect the hosts' orbital reflex motion, even for state-of-the-art infrared Doppler spectrometers on 10-meter-class telescopes (Mahadevan et al. 2014). In the case of transit surveys, this is due to their especially faint optical-wavelength magnitudes, where transit surveys have historically operated. For those with measured optical magnitudes, none have a *g*-band magnitude brighter than 21, making it prohibitively difficult to achieve the signal-to-noise required to detect a transiting satellites at these wavelengths ($\lambda \sim 0.5\ \mu$m), even with meter-class space-based telescopes like *Kepler*. Figure 3 shows all transiting planets as of the submission of this white paper, as a function of host star mass and orbital period (left) and effective temperature and orbital semi-major axis (right). Current state-of-the-art transiting planet surveys, such as *Kepler* (Borucki et al. 2009), K2 (Howell et al. 2014), *TESS* (Ricker et al. 2015), KELT (Gaudi et al. 2017), WASP (Pollacco et al. 2006), HAT (Bakos et al. 2004), MEarth (Charbonneau et al. 2009), TRAPPIST (Gillon et al. 2013), and SPECULOOS (Burdanov et al. 2017) all operate at wavelengths less than 1.1 $\mu$m, making them sensitive to transiting planets orbiting F, G, K, M and some early L-type stars. However, most L and T dwarfs have maximum flux values at infrared wavelengths (1.1 to 2.5 $\mu$m), and the brightest 450 L and T dwarfs all have *J*-band ($\lambda \sim 1.25\ \mu$m) magnitudes brighter than 16. If operated at infrared wavelengths, a transit survey is likely to succeed at discovering new exosatellites.

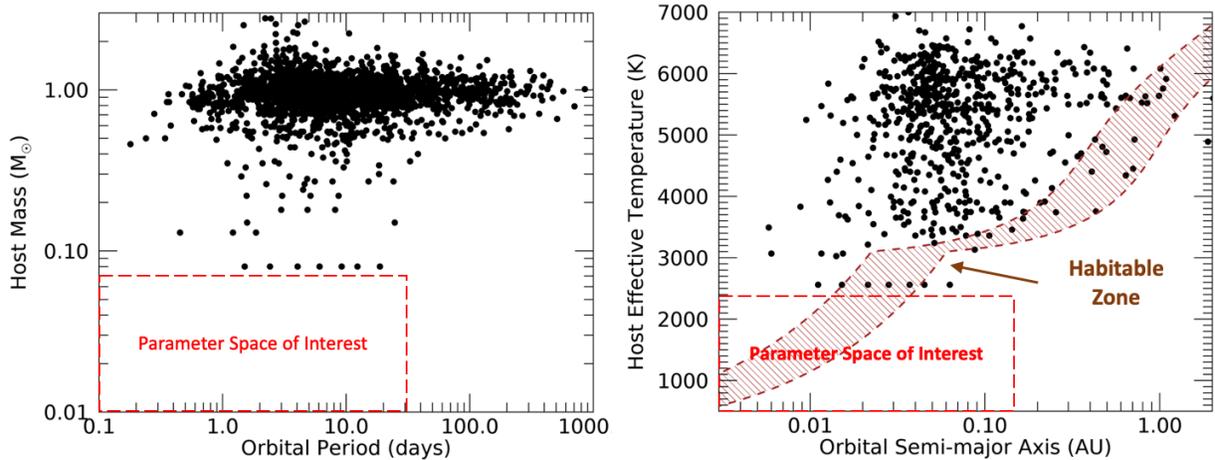

**Figure 3:** Host mass versus orbital period (left) and host effective temperature versus orbital semi-major axis (right) for confirmed transiting exoplanets in the NASA Exoplanet Archive. NASA's *Kepler*, K2, and *TESS* Missions operate at optical wavelengths and are optimized for sun-like and early M dwarf stars. The ground-based MEarth survey (Charbonneau et al. 2009) and TRAPPIST survey (Gillon et al. 2016) operate at red optical wavelengths to effectively survey mid-to-late M dwarfs. In this white paper, we argue that surveys at *infrared* wavelengths with robotic 1-2-meter telescopes will discover short-period transiting satellites orbiting L and T dwarfs (the "parameter space of interest" in each figure). The habitable zone is indicated in the right-hand plot, calculated from Kopparapu et al. 2013 and extrapolated for effective temperatures below 2600 K. The discontinuity in the habitable zone near 3000 K arises from a change in the stellar radius-$T_{eff}$ relation for LT dwarfs.

**Searching for Transiting Satellites Orbiting L and T Dwarfs in the 2020s: A Natural Fit for the US 1-2-meter Optical/Infrared Fleet at Minimal Cost**

Recent advances and industrialization of telescope robotization and near-infrared detector technologies present the the opportunity to efficiently search for satellites that transit L and T dwarfs with the current fleet of US 1-to-2-meter class optical/infrared (O/IR) telescopes. Historically, the high cost of large-format infrared detectors have limited their deployment to large telescopes with diameters greater than 2-meters. Recently, low-cost large-format InGaAs detectors have appeared on the commercial-off-the-shelf (COTS) market, and academic investigations show they can meet state-of-the-art infrared detector performance (e.g., Sullivan et al. 2013).

Furthermore, the demand for retrofitting and robotization of 1-2-meter telescopes from the high-end amateur market has led to a myriad of companies that sell commercial telescope and dome control systems designed for automated observations. The standardization of telescope, dome and device driver communications enables rapid implementation of automated operations (e.g., Astronomy Common Object Model, or ASCOM, Denny 2002). The combination of low-cost infrared detector technology and low-cost robotization makes the 1-2-meter US O/IR fleet well-positioned to execute a search for transiting satellites orbiting L and T dwarfs.  Lastly, bright moonlight does not significantly impact infrared observations, and infrared programs can operate during bright time without impacting on-going dark time programs with 1-2-meter telescopes.

**Ancillary Science from Robotic Infrared LT Monitoring: Variability of L and T Dwarfs**
A monitoring campaign in search of satellites around L and T dwarfs will produce a wealth of light curve data from which intrinsic brightness variations of ultracool dwarfs and their respective mechanisms can be studied. Variations of objects at the high end of the L-T temperature range can further our understanding of electrical conductivity, and magnetically driven phenomena in cool stellar atmospheres, including spots and auroral processes that may be induced by orbiting satellites as in the Jupiter-Io system (Lane et al. 2007, Hallinan et al. 2015).  At lower temperatures, variations are most likely related to the formation and dissipation of dust clouds, and complex weather patterns, although there remains debate over the specific processes involved (e.g., Radigan et al. 2014, Metchev et al. 2015). These lower temperature brown dwarfs play an especially important role in our understanding of condensates in substellar atmospheres, with light curves providing unique constraints on the heights, surface distributions, and dynamics of cloud features, for which similar planetary data is unavailable (e.g., Apai et al. 2013, 2017).

**Ancillary Science from Robotic Infrared LT Monitoring: Parallax Observations**
Many currently known L and T dwarfs are nearby, enabling precise distances to be found using ground-based, infrared parallax measurements (Faherty et al. 2012, Dupuy et al. 2012).  Long-term, ground-based photometric monitoring of L and T dwarfs will provide many epochs of data that can be used for precise astrometry (see e.g., Dittmann et al. 2014).  While the *Gaia* mission has revolutionized our understanding of the Galaxy for objects at earlier spectral types, most L and T dwarfs are too red and too faint for astrometry from *Gaia*.  Cross-matching L dwarfs from the Sloan Digital Sky Survey with *Gaia* DR2, we find that only ~50% of L0–L2 dwarfs have an astrometric solution from *Gaia* DR2 with a sharp decline toward later spectral types (Theissen 2018; Kirkpatrick et al. 2019; Smart et al. 2019; Skinner et al. in prep). Ground-based monitoring of L and T dwarfs would offer a powerful extension to *Gaia* and ensure that satellite-hosting L and T dwarfs had accurate distances and luminosities.  In addition to enabling detailed characterization of satellites orbiting L and T dwarfs, precise distances to a large sample of L and T dwarfs will enhance our understanding of the lowest-mass end of the luminosity function of our Galaxy.

**Ancillary Science from Robotic Infrared LT Monitoring: Target-of-Opportunity Interrupts**
Robotization of the 1-2-meter OIR fleet and infrared operations during bright time naturally enable automated target-of-opportunity interrupt observations in support of transient and multi-messenger astrophysics projects.